\documentclass[pdflatex,sn-mathphys-num]{sn-jnl}

\usepackage{amsmath, amsthm, amssymb, amsfonts, enumerate}
\usepackage[dvipsnames]{xcolor}
\usepackage{lmodern,microtype,ragged2e, comment}
\usepackage{appendix,hyphenat,mathtools,dsfont,bm,bbm}
\usepackage[makeroom]{cancel}
\usepackage{graphicx}
\usepackage{soul}
\usepackage[normalem]{ulem}
\usepackage{hyperref}
\usepackage{ifthen}
\usepackage{xspace}
\usepackage[caption=false]{subfig}
\usepackage{physics}
\usepackage{float}
\usepackage{booktabs}
\usepackage{algorithm}%
\usepackage{algorithmicx}%
\usepackage{algpseudocode}%
\usepackage[T1]{fontenc} 
\usepackage[utf8]{inputenc} 

\newtheorem{theorem}{Theorem}

\newcommand{\proj}[1]{\ket{#1}\!\bra{#1}}

\newcommand{\bematrix}{\left(\begin{matrix}}
\newcommand{\ematrix}{\end{matrix}\right)}
\def\one{{\mbox{$1 \hspace{-1.0mm}  {\bf l}$}}}						

\def\C{{\ensuremath{\mathbbm C}}}
\def\R{{\ensuremath{\mathbbm R}}}

\def\ii{\mathrm{i}}
\def\tr{\mathrm{tr}}

\def\cB{\mathcal B}

\def\cH{\mathcal H}

\def\cO{\mathcal O}

\def\cX{\mathcal X}

\newcommand\defn[1]{\textsl{#1}}

\newcommand{\eqnref}[1]{Eq.~(\ref{#1})}

\newcommand{\figref}[1]{Fig.~\ref{#1}}

\newcommand{\secref}[1]{Sec.~\ref{#1}}

\newcommand{\beq}{\begin{equation}}
\newcommand{\eeq}{\end{equation}}

\usepackage{tcolorbox} 
\usepackage{amsthm} 
\usepackage{algpseudocode} 
\newtheorem{definicion}{Definition}[section]

\begin{document}

\title{Entanglement Detection with Quantum-inspired Kernels and SVMs}

\author[1]{\fnm{Ana} \sur{Martínez-Sabiote}}\email{anamarsabi@ugr.es}

\author[1,2]{\fnm{Michalis} \sur{Skotiniotis}}\email{mskotiniotis@onsager.ugr.es}

\author[1,2]{\fnm{Jara J.} \sur{Bermejo-Vega}}\email{jbermejovega@onsager.ugr.es}

\author[1,2]{\fnm{Daniel} \sur{Manzano}}\email{manzano@onsager.ugr.es}

\author[3]{\fnm{Carlos} \sur{Cano}}\email{carloscano@ugr.es}

\affil[1]{\orgdiv{Electromagnetism and Matter Physics Department}, \orgname{University of Granada}, \orgaddress{\street{Avenida de Fuentenueva S/N}, \city{Granada}, \postcode{18071}, \state{Granada}, \country{Spain}}}

\affil[2]{\orgdiv{Institute Carlos I of Theoretical and Computational Physics}, \orgname{University of Granada}, \orgaddress{\street{Avenida de Fuentenueva S/N}, \city{Granada}, \postcode{18071}, \state{Granada}, \country{Spain}}}

\affil[3]{\orgdiv{Department of Computer Science and A.I.}, \orgname{University of Granada}, \orgaddress{\street{Calle Periodista Daniel Saucedo Aranda S/N}, \city{Granada}, \postcode{18071}, \state{Granada}, \country{Spain}}}

\abstract{
This work presents a machine learning approach based on support vector machines (SVMs) for quantum entanglement detection. Particularly, we focus in bipartite systems of dimensions $3\times 3$, $4\times 4$, and $5\times 5$, where the positive partial transpose criterion (PPT) provides only partial characterization.
Using SVMs with quantum-inspired kernels we develop a classification scheme that distinguishes between separable states, PPT-detectable entangled states, and entangled states that evade PPT detection. Our method achieves increasing accuracy with system dimension, reaching $80\%$, $90\%$, and nearly $100\%$ for $3\times 3$, $4\times 4$, and $5\times 5$ systems, respectively. Our results show that principal component analysis significantly enhances performance for small training sets. The study reveals important practical considerations regarding purity biases in the generation of data for this problem and examines the challenges of implementing these techniques on near-term quantum hardware. Our results establish machine learning as a powerful complement to traditional entanglement detection methods, particularly for higher-dimensional systems where conventional approaches become inadequate. The findings highlight key directions for future research, including hybrid quantum-classical implementations and improved data generation protocols to overcome current limitations. 
}


\maketitle

\section{Introduction}

Quantum computing and machine learning are rapidly growing fields poised for cross-fertilization. First, quantum computers can beat their classical counterparts for many tasks, making them potentially useful for machine learning~\cite{manzano:njp09,dunjko:rpp18,schuld2021machine,Dunjko2016}. Examples of {\it quantum-enhanced} machine learning algorithms are Variational Quantum Algorithms~\cite{Cerezo2021}, Quantum Neural Networks~\cite{killoran:prr19,torres:njp22,Torres2024}, and quantum walk-based algorithms~\cite{Paparo2014,marin:2025}. These techniques  have already been applied to Reinforcement Learning~\cite{cuellar2024automatic}, Time Series analysis~\cite{cuellar2023time} and many others Machine Learning problems~\cite{biamonte2017quantum}. Furthermore, quantum computing involves many computationally hard problems that can be solved by autonomous algorithms. Examples are the design of quantum experiments \cite{Melnikov2018}, as well as tomography of quantum states \cite{Cha2022,Palmieri2020} and processes \cite{Torlai2023}. 

Entanglement stands as one of the most profound and defining features of quantum mechanics. Originally introduced by Einstein, Podolsky and Rosen as part of their argument against the completeness of the theory \cite{einstein:pr35}, it was later identified by Schrödinger as the key defining feature of quantum mechanics \cite{schrodinger:mpcps35}. Beyond its foundational and philosophical significance, entanglement plays a vital role in the advancement of quantum technologies. It underpins a range of pivotal techniques such as quantum teleportation \cite{bennet:prl92,pirandola:np15}, measurement-based quantum computation  \cite{raussendorf:pra03,briegel:np09}, and cryptography \cite{Ekert1991,Yin2020}. A central challenge related to entanglement is the development of criteria for separability and measures of entanglement \cite{horodecki:rmp09}. This involves determining whether a given quantum state exhibits entanglement. Several approaches have been proposed, including the renowned Bell inequalities \cite{bell:p64,clauser:prl69}, the Peres-Horodecki positive partial transpose (PPT) criterion \cite{peres:prl96,horodecki:pla96}, as well as entropic criteria \cite{plastino:epl09,manzano:jpa10}. For quantum systems with dimensions up to $2\times 3$, the PPT criterion provides both necessary and sufficient conditions for detecting entanglement in pure or mixed states. However, for higher-dimensional systems, no general sufficient conditions are currently known.

Due to the importance of this problem, in the last years several solutions based on machine learning have been proposed. These include solutions based on classical neural networks \cite{Urena2024,Asif2023}, variational quantum algorithms \cite{Wang2022,Munoz-Moller2022}. A recent approach is also based on Support Vector Machines (SVMs), or kernel methods, both classical \cite{casale2023large} and quantum \cite{Mahdian2025}. In this paper, we follow this line and develop separability algorithms based on support vector machines for systems of dimensions $3 \times 3$, $4 \times 4$, and $5 \times 5$, all of them beyond the detection capability of the PPT criterion. As in Ref. \cite{casale2023large}, we separate the entangled states in two classes, those that can be detected by the PPT criterion, and those that cannot. We analyze the role of the Principal Component Analysis (PCA) technique showing that it plays a major role in the capability and simulability of the model. Finally, we also study biases in the data generation as well as potential applications of quantum methods to handle this same problem. 

This paper is organized as follows. In Section \ref{sec:background}, we introduce the fundamental concepts necessary to understand both the problem at hand and the machine learning techniques employed. We begin with an overview of entanglement and entanglement detection, followed by a description of SVMs and kernel methods. Section \ref{sec:methods} outlines the methodology, including the generation of quantum datasets and the implementation of the SVMs. In Section \ref{sec:results}, we present and analyze the results of our simulations, with a particular focus on comparing different kernel functions and assessing the impact of Principal Component Analysis (PCA). Finally, in Section \ref{sec:discussion}, we address several key aspects, such as the influence of potential biases in data generation and the challenges of deploying the approach on real quantum hardware. The paper concludes with a summary of our main findings.

\section{\label{sec:background}Background}
In this section we review the necessary background underpinning the findings of our work. In \secref{sec:QIT} we 
review the key elements of quantum information theory, and introduce the separability 
problem, while in \secref{sec:SVM} we explain the basic theory behind support vector machines and introduce our quantum inspired kernel function.

\subsection{\label{sec:QIT}Quantum Information Theory}

Here we introduce the basic concepts of quantum information theory.  We limit ourselves to only those concepts that 
are necessary to follow the rest of this article. The interested reader can find a more thorough and complete 
exposition of quantum information theory in Refs~\cite{Nielsen_Chuang_2010, Watrous_2018}.

All information pertaining to the properties of a quantum system is contained in its quantum state. The latter is an 
element of a complex Euclidean space---henceforth referred to as a \defn{Hilbert} space and denoted as $\cH$---that 
is complete in its norm.  Throughout this work we shall be interested in finite dimensional Hilbert spaces, i.e., 
$\mathrm{dim}(\cH)=d\leq \infty$. The state of a quantum system is mathematically described by a $d$-dimensional
vector $\ket{\psi}\in\cH$, where $\ket{\cdot}$ is the customary Dirac notation for a column vector.  
Row vectors---denoted by $\bra{\cdot}$---correspond to linear functionals, i.e., $\bra{\cdot}:\cH\to\C$.  
Defining an orthonormal basis, $\{\ket{i}\, \vert\, i\in(0,\ldots,d-1\}\in\cH$, any state vector $\ket{\psi}\in\cH$ 
can be expressed with respect to this basis as
    \begin{equation}
        \ket{\psi} = \sum_{i=0}^{d-1}\, \psi_i\, \ket{i}\, ,
        \label{eq:vecexpansion}
    \end{equation}
where $\psi_i=\braket{i}{\psi}\in\C$, represents the inner product between vectors $\ket{i}$ and $\ket{\psi}$.

An alternative yet equivalent way of describing the state of a quantum system is the \defn{density operator} 
formalism.  A density operator $\rho:\cH\to\cH$ is a linear, positive semi-definite operator of unit trace 
acting on the state space $\cH$. The set of all bounded linear operators acting on $\cH$---denoted by $\mathcal{B}(\cH)$---is 
itself a vector space equipped with the inner product defined by the trace, $\tr[\cdot]$, 
\begin{equation}\label{eq:HilbertSchmidt}
(A,B)=\tr[A^\dagger B], \forall\, A,B\in\mathcal{B}(\cH)\,,
\end{equation}
known as the {\it Hilbert-Schmidt} inner product. It follows that there exists an orthonormal basis 
$\{\ketbra{i}{j}\, \vert \, i,j\in(0,\ldots,d-1)\}$ such that 
    \begin{equation}
        \rho = \sum_{i,j=0}^{d-1} \rho_{ij}\,\ketbra{i}{j}\, ,
        \label{eq:rhoexpansion}
    \end{equation}
where $\rho_{ij}=\tr[\ketbra{j}{i}\rho]\in\C$. However, as $\rho\in\mathcal{B}(\cH)$ is positive semi-definite it is also 
Hermitian, i.e., $\rho=\rho^\dagger$ where $\dagger$ denotes the conjugate transpose operation. 
It is thus convenient to expand $\rho$ with respect to a basis of Hermitian operators.  For any dimension $d$
one can always find a set of $d^2-1$ Hermitian and traceless matrices 
$\{G_i\, \vert\, G_i=G^\dagger_i, \, \tr(G_i)=0 \,\,\forall\, i\in(1,\ldots, d^2-1)\}$ which, along with the 
$d$-dimensional identity matrix $\one$, form an orthogonal basis for $\mathcal{B}(\cH)$.  With respect to this basis 
$\rho\in\mathcal{B}(\cH)$ can be written as    
    \begin{equation}
        \rho = \frac{\one + {\bf r}\cdot {\bf G}}{d}\, 
    \label{eq:Blochrep}
    \end{equation}
where ${\bf r}\in\R^{d^2-1}$, $0\leq \abs{\bf r}\leq \sqrt{d-1}$, and ${\bf G} = (G_1,\ldots, G_{d^2-1})^T$~\footnote{Note 
that the denominator is necessary in order to ensure that $\tr(\rho)=1$.}.  A quantum system in some state 
$\ket{\psi}\in\cH$ has corresponding density operator given by $\rho=\ketbra{\psi}{\psi}$.

The advantage of the density matrix formalism is that it provides a more general description of the state of a 
quantum system.  Consider such a system known to be in one of a set of quantum state 
$\{\ket{\psi_i}\,\vert\, i\in\mathbb{Z}\}$ with corresponding probability $p_i$. The \defn{ensemble} 
$\{p_i, \, \ket{\psi_i}\, \vert \, i\in \mathbb{Z}\}$ is described by the density operator 
    \begin{equation}
        \rho = \sum_{i}p_i\, \ketbra{\psi_i}{\psi_i}\, .
        \label{eq:ensemble}
    \end{equation}
Notice that whilst a given ensemble gives rise to a unique density matrix the converse is not true.  Indeed, for a 
given $\rho\in\mathcal{B}(\cH)$ there correspond an infinitude of possible ensembles of quantum states. One such ensemble is 
furnished by the spectral decomposition of $\rho\in\mathcal{B}(\cH)$
    \begin{equation}
        \rho = \sum_{k=1}^r \lambda_k\, \ketbra{\lambda_k}{\lambda_k}\,,
        \label{eq:spectraldecomp}
    \end{equation}
where $0<\lambda_k\leq 1, \, \sum_{k=1}^r \lambda_k = 1$ are the non-zero eigenvalues of $\rho$ 
(including possible repetitions) and $\{\ket{\lambda_k}\, \vert \braket{\lambda_k}{\lambda_\ell}=\delta_{k\ell}, 
k,\ell\in(1,\ldots,r)\}$ are the corresponding eigenvectors. The number of non-zero eigenvalues of $\rho\in\mathcal{B}(\cH)$, 
$r$, is known as its \defn{rank}.  Rank-one density operators, i.e., $\rho=\ketbra{\psi}{\psi}$, are referred to as 
\defn{pure} quantum states, whereas density operators with $r>1$ are referred to as \defn{mixed} quantum states.
It can be verified that $\abs{\bf r}$ in \eqnref{eq:Blochrep} attains its maximum value if and only if 
$\rho\in\mathcal{B}(\cH)$ is pure. 
Equivalently, $\tr(\rho^2)=1$ if and only if $\rho\in\mathcal{B}(\cH)$ is pure. The magnitude of ${\bf r}$
can be directly related to the purity of $\rho$ as $\abs{\bf r}^2=d\,\tr\rho^2-1$. 
We will often use the much simpler state vector formalism $\ket{\psi}\in\cH$ to denote pure quantum states 
and the density operator $\rho\in\mathcal{B}(\cH)$ to denote mixed quantum states.

The state of a quantum system can be altered by means of a \defn{quantum operation}.  The latter is described by 
a map $\Phi:\mathcal{B}(\cH)\to\cB(\cH'), \; \rho\xrightarrow{\Phi}\rho'\in\cB(\cH')$ that is linear, trace non-increasing, 
and completely positive (for a more detailed discussion on quantum operations see~\cite{Nielsen_Chuang_2010, 
Watrous_2018}).  Quantum operations provide the most general description on how the state of a quantum system can be 
transformed including unitary transformations, measurements, quantum state preparation as well as encoding of 
classical information into quantum states. Let $\mathcal{X}\subset \R^n$ denote a classical alphabet we wish to 
encode onto the state of a quantum system. Then the \defn{encoding map} $\Phi$
\begin{equation}
    \Phi: \mathcal{X}\to\mathcal{B}(\cH)
    \label{eq:encodingmap}
\end{equation}
describes the quantum state preparation procedure 
\begin{equation}
\bf x\xrightarrow{\Phi}\rho(\bf x)\in\mathcal{B}(\cH), \, \forall\, \bf x\in\mathcal{X}\, .
    \label{eq:statepreparation}
\end{equation}
%
The final ingredient we need is how to compose the state spaces of multiple systems.  Consider a composite 
system consisting of $N$ constituent quantum systems with corresponding Hilbert spaces $\cH_n,\; n\in(1,\ldots,N)$.
The Hilbert space corresponding to the composite system is the \defn{tensor product} of each of the Hilbert 
spaces of the constituent parts, i.e.,
    \begin{equation}
        \cH = \cH_1\otimes\cH_2\otimes\ldots\otimes\cH_N:=\bigotimes_{n=1}^N \cH_{n}\,.
    \label{eq:tensorproduct}
    \end{equation}
Throughout this article we shall be concerned with \defn{bipartite} quantum systems---quantum systems with only two subsystems of interest---whose state spaces we shall denote by $\cH_A$ and $\cH_B$.  A well known theorem by Schmidt allows 
one to express any pure state $\ket{\psi}\in\cH_A\otimes\cH_B$ in a standard form.
    \begin{theorem}[Schmidt decomposition]
        Let $\ket{\psi}\in\cH_A\otimes\cH_B$ where $\mathrm{dim}(\cH_A)=d_A\neq d_B=\mathrm{dim}(\cH_B)$.  There 
        exists a pair of orthonormal basis $\{\ket{u_i}_A\}$ and $\{\ket{v_j}_B\}$ such that 
            \begin{equation}
                \ket{\psi} = \sum_{k=1}^r \sqrt{\lambda_k}\, \ket{u_k}_A\otimes\ket{v_k}_B\, ,
            \label{eq:Schmidt_decomp}
            \end{equation}
        where $0<\lambda_k\leq 1, \; \sum_{k=1}^r\lambda_k=1$ are the \defn{Schmidt coefficients} of the state 
        $\ket{\psi}\in\cH_A\otimes\cH_B$ and $r\leq\min\{d_A,d_B\}$ is the states \defn{Schmidt rank}.     
    \end{theorem}

A proof of this theorem can be found in~\cite{Nielsen_Chuang_2010}. A bipartite state 
$\ket{\psi}\in\cH_A\otimes\cH_B$ is said to be of \defn{product form}, i.e., 
    \begin{equation}
        \ket{\psi} = \ket{u}_A\otimes\ket{v}_B\, ,
    \end{equation}
if and only if it has Schmidt rank equal to unity.  Pure bipartite states with Schmidt rank greater than one are 
said to be \defn{entangled}. For pure bipartite states the Schmidt decomposition also allows one to quantify the 
amount of entanglement present in a quantum state.  An important result due to Nielsen imposes a 
total ordering of all bipartite quantum states according to their entanglement content~\cite{Nielsen99}. For systems composed of more 
than two parts, however, no such ordering exists. Indeed for tripartite systems pure quantum states can be entangled 
in two inequivalent ways~\cite{Dur00}, whilst the number of inequivalent classes of entangled states proliferates the 
more systems are considered~\cite{Spee16}. 

The question of whether mixed bipartite states are entangled or not is much more subtle as we now explain.  Let 
$\rho\in\cB(\cH_A\otimes\cH_B)$.  The separability problem consists of determining whether $\rho$ admits a decomposition of the form
\begin{equation}
    \rho = \sum_{j=1}^k q_j \rho_{A,j} \otimes \rho_{B,j} \text{ such that } q_j\in \R^{+},\, \sum_{j=1}^k q_j =1 \, ,
\label{eq:sep_state}
\end{equation}
where $\{\, q_j, \, \rho_{A,j} \otimes \rho_{B,j}\}$ is some ensemble of $\rho$ with $\rho_{A,j}\in\cB(\cH_A), \forall\, j$ and $\rho_{B,j}\in\cB(\cH_B), \forall\, j$. If $\rho$ admits such a decomposition, then it is said 
to be separable, otherwise it is said to be entangled. The difficulty of the separability problem stems from the fact that one needs to 
check over an infinitude of decompositions of $\rho\in\cB(\cH_A\otimes\cH_B)$ in order to determine whether it is entangled or not.
For bipartite systems with dimensions $d_A=d_B=2$ or $d_A=2, d_B=3$ a necessary and sufficient condition for separability is the  
Peres-Horodecki, or Positive Partial Transpose (PPT) criterion~\cite{Peres96,Horodecki96}.

\begin{definicion}[Peres-Horodecki Criterion]
    Let $\mathcal{H} = \mathcal{H}_A \otimes \mathcal{H}_B$ be a bipartite Hilbert space, $\rho \in \mathcal{B}(\mathcal{H})$ a density operator and let $\rho^{T_B}$ denote the partially transposed density operator with respect to $\mathcal{H}_B$. Then $\rho$ fulfills the Peres-Horodecki criterion with respect to the bipartition $\mathcal{H} = \mathcal{H}_A \otimes \mathcal{H}_B$ if and only if $\rho^{T_B}$ (or equivalently $\rho^{T_A}$) has only non-negative eigenvalues. 
\end{definicion}

Generally speaking, the separability problem for mixed bipartite states of dimension $d_A, d_B>3$ is known to be 
NP-hard~\cite{Gurvits2003, Horodecki2009, Gutoski2013, Hiesmayr2021}. In these cases the Peres-Horodecki criterion is a necessary 
but not sufficient condition for separability.  Using the Peres-Horodecki criterion all mixed bipartite states can be classified into one 
of the following three classes of states:
\begin{description}
    \item[\textbf{Separable}] density matrices that can be decomposed as specified in \eqnref{eq:sep_state} and, therefore, fulfill the Peres-Horodecki criterion  
    \item[\textbf{PPT entangled}] entangled density matrices that fulfill the PPT criterion. 
    \item[\textbf{NPPT entangled}] entangled density matrices that do not fulfill the PPT criterion.
\end{description}

Thus given a classical description---as provided by the $(d^2-1)$-dimensional vector ${\bf r}$ in \eqnref{eq:Blochrep}---of a bipartite 
mixed quantum state our aim is to determine in which of the three afore mentioned classes of states it belongs. Such classification 
problems are the bread and butter of supervised learning algorithms which we now briefly introduce. 

\subsection{\label{sec:SVM}Support Vector Machines}

SVMs are a class of supervised learning algorithms widely used for classification 
tasks. Here we introduce the basic concepts, but readers are referred to~\cite{Scholkopf2001, Combarro2023} for more detailed descriptions of classical and quantum SVMs. 

At the supervision stage, SVMs are trained using a set of data $\{ (\mathbf{x}_i, y_i) \}_{i=1}^{n}$, 
where $\mathbf{x}_i \in \R^d$ represent the pattern vectors and $y_i \in \{ -1, 1 \}$ are the corresponding labels
for their classes (henceforth we shall assume a binary classification problem). The goal of SVMs is to find a hyperplane in 
$\R^D$ (typically with $D>d$) that separates data points from both classes by maximizing the margin distance between the 
nearest points (support vectors) from each class. The training of an SVM can be thus posed as the following optimization problem 
    \begin{equation}\label{eq:SVM_hardMargin}
        \text{Minimize} \quad \frac{1}{2} \|\mathbf{w}\|^2, \quad \text{subject to} \quad y_i (\mathbf{w} \cdot \mathbf{x}_i + b) \geq 1, 
        \quad \forall i \in [1,n],\, 
    \end{equation}
where the hyperplane is defined by the normal vector $\mathbf{w} \in \R^D$ and a constant $b \in \R$. The above 
formulation is called the {\it hard-margin} SVM, as it does not allow for any classification errors. However, if the data is not 
linearly separable, the training can consider adjustable slack coefficients $\xi_i \geq 0$ to permit some degree of misclassification. 
In this case, the optimization problem becomes
    \begin{equation}\label{eq:SVM_softMargin}
        \text{Minimize} \quad \frac{1}{2} \|\mathbf{w}\|^2 + C \sum_{i=1}^{n} \xi_i, \quad \text{subject to}, 
        \quad y_i (\mathbf{w} \cdot \mathbf{x}_i + b) \geq 1 - \xi_i, \quad \forall i \in [1,n], \xi_i \geq 0,\,
    \end{equation}
where $C \geq 0$ is a regularization parameter controlling the trade-off between maximizing the margin and minimizing classification 
errors. This formulation of the task is called the {\it soft-margin} SVM and can be equivalently cast in its dual formulation as 
    \begin{equation}\label{eq:SVM_lagDualForm}
        \text{Maximize} \quad \sum_{i=1}^{n} \alpha_i - \frac{1}{2} \sum_{i,j}^{n} y_i y_j \alpha_i \alpha_j (\mathbf{x}_i \cdot 
        \mathbf{x}_j), \quad \text{subject to}, \quad 0\leq\alpha_i\leq C \quad \sum_{i}^{n} \alpha_i y_i = 0.
    \end{equation}    
Therefore, training a SVM amounts to solving a very large quadratic programming optimization problem for finding the values of  
$\alpha_i (i\in[1,n])$ satisfying \eqnref{eq:SVM_lagDualForm} for which there exist well-known classical methods yielding an approximate 
solution~\cite{platt:svm98}. Once a solution has been found, the input pattern vectors $\mathbf{x}_{i}$ and their corresponding 
$\alpha_i >0$ are the support vectors from which one can recover $\mathbf{w}$ as 
    \begin{equation}\label{eq:w_vec}
        \mathbf{w}=\sum_{i}^{n} \alpha_i\, y_i \,\mathbf{x}_i. 
    \end{equation}
In order to classify a new point, $\mathbf{x}_{new}\in\R^d$ we simply compute 
    \begin{equation}\label{eq:SVM_lagDualFormPredict}    
        \mathbf{w}\cdot \mathbf{x}_{new} + b = \sum_{i}^{n}\alpha_i y_i (\mathbf{x}_i\cdot \mathbf{x}_{new}) +b
    \end{equation}
and check whether the result is positive or negative. Note that in the dual formulation of the soft-margin SVM problem, 
\eqnref{eq:SVM_lagDualForm}, it is clear that all the information one needs from the input pattern vectors $\mathbf{x}_{i}$ are their 
inner products. Indeed, we can express both the training, \eqnref{eq:SVM_lagDualForm}, and the predictive phase, 
\eqnref{eq:SVM_lagDualFormPredict}, of a SVM purely using the inner products of the pattern vectors.

In order to apply SVMs to real-world scenarios, where the data is not linearly separable, we first need to represent the patterns  
$\mathbf{x}_i \in \cX\subset\R^d$ as vectors in some inner product space $\mathcal{V}$ called a \emph{feature space} which need not coincide 
with $\R^d$. This representation is achieved by an encoding map (also known as the \emph{feature map}), 
$\Phi : \mathcal{X} \longrightarrow \mathcal{V}$.  Calculating the mapping $\Phi$ explicitly can be computationally expensive for high-dimensional spaces. 
SVMs cleverly circumvent this cost by making use of the so-called \emph{kernel trick}.  A \emph{kernel} is a linear bi-variate 
real valued function $\kappa : \mathcal{X} \times \mathcal{X} \longrightarrow \R$.  Using such a kernel an SVM exploits the following fact   
 \begin{equation}
     \kappa(\mathbf{x}, \mathbf{y}) = \left< \Phi(\mathbf{x}), \Phi(\mathbf{y}) \right>_{\mathcal{V}}, \quad \forall\, \mathbf{x},\mathbf{y}\in\cX,
 \end{equation}
in order to operate in high-dimensional spaces. Commonly used classical kernels include linear, polynomial or Radial Basis Function (RBF) kernels, among others (see~\cite{Scholkopf2001}). 

Quantum kernels naturally arise when the feature space is a $d$-dimensional Hilbert space and the encoding map is a general quantum 
operation~\cite{schuld:qml19, schuld2021machine}. Specifically, given a feature map represented by the quantum operation $\Phi: \mathcal{X}\to\cB(\cH)$, 
the corresponding quantum kernel is the Hilbert-Schmidt inner product, \eqnref{eq:HilbertSchmidt}, between the quantum states $\rho(\mathbf{x}), \rho(\mathbf{y})$ corresponding to the feature vectors $\mathbf{x},\mathbf{y} \in \mathcal{X}$, i.e.,
    \begin{equation}
        \kappa(\mathbf{x},\mathbf{y}) = \tr\left[ \rho(\mathbf{x})\rho(\mathbf{y})\right].
    \end{equation}
For instance, the encoding map may correspond to a quantum circuit that maps classical data into quantum states. Note that quantum feature maps, and their associated quantum circuits, will give rise to different quantum kernels~\cite{schuld2021machine}.

\section{\label{sec:methods} Methods}

In this section we describe how we generate data in the form of quantum states for our classification task in \secref{sec:dataset}, 
whilst \secref{sec:Implementation} describes the implementation of both classical as well as quantum inspired SVMs.     

\subsection{\label{sec:dataset}Data Generation}

To train and evaluate SVMs to tackle the separability problem we first need a dataset with descriptions of entangled 
(both PPT and non-PPT) and separable bipartite quantum systems. A classical description of these density matrices is provided via the 
state vector $\bf r$ as discussed in \secref{sec:background}. Following the approach of Ref~\cite{casale2023large}, to  generate $d\times d$ density matrices we proceed as follows:

\begin{algorithm}
\begin{algorithmic}[1]
\State Generate a random $d\times d$ complex matrix $A$;
\State Normalize $A$ resulting in  $\rho=A^{\dagger} A/\Tr \left( A^{\dagger} A\right)$. 
\newline
\textbf{return} $\rho$
\end{algorithmic}
\end{algorithm}

The generation of separable density matrices proceeds straightforwardly from the definition of \eqnref{eq:sep_state}. For state spaces 
$\cH_A, \cH_B$, of dimension $d_A, d_B$ respectively, we generate sets of $d_A\times d_A$ and $d_B\times d_B$ matrices.  We then pick 
pairs of these density matrices according to some probability distribution $\{q_j\}$, compose them via the tensor product, and sum them.  
Entangled NPPT quantum states are generated by sampling density matrices of dimension $d_Ad_B\times d_Ad_B$ and keeping only those that do 
not fulfill the PPT criteria.




More effort is dedicated to the generation of entangled PPT quantum states, as for this case there are no clear separability criteria. Therefore, we first sample random quantum states and keep those that respect the PPT condition. After that we determine if they are entangled by computing 1000 iterations of a variation of the Frank Wolfe algorithm \cite{frank56} to find its nearest separable quantum state. This algorithm is based in a minimisation performed over a specific set of matrices, the separable set in our case. To apply it to a density matrix $\rho$, the Frank Wolfe algorithm starts from a random candidate state $\rho_0$. First, it finds the largest eigenvector of the matrix $(\rho -\rho_0)$, $\ket{s^*}$ . This eigenvector approximates $\rho$, it is easy to calculate without full-diagonalisation, but it is not necessarily separable. To restrict our optimization to the separable states, we calculate the Schmidt decomposition of this state $\ket{s^*}=\sum_i \lambda_i \ket{A}_i \otimes \ket{B}_i$, then we select the state $\rho_1=\ket{A}_1 \otimes \ket{B}_1$ corresponding to the highest Schmidt coefficient as our new candidate state. By iterating, we find a close separable state to our original one. Then, we select those states whose distance to their separable  approximation is bigger than a certain threshold (set to $0,01$). This way we obtain a set of states that fulfill the PPT condition but are very likely to be entangled. The following pseudocode summarises the Frank-Wolfe method:

\begin{algorithm}
\caption{Frank-Wolfe algorithm}\label{alg:Frank-Wolfe}
\begin{algorithmic}[1]
\State $\rho_0 \gets \ket{\psi_A}\bra{\psi_A} \otimes \ket{\psi_B}\bra{\psi_B}$;
\For{$t\gets 0 \textbf{ to } T$}   
    \State  $\ket{s^*} \gets \text{largest eigenvector of }(\rho -\rho_t)$;
    \State  $\ket{s^*_A}, \ket{s^*_B} \gets \text{largest Schmidt states of} \ket{s^*}$:
    \State  $\alpha \gets \frac{2}{t+2}$;
    \State  $\rho_{t+1} \gets (1-\alpha)\rho_t + \alpha \ket{s^*_A}\bra{s^*_A} \otimes \ket{s^*_B}\bra{s^*_B}$;
\EndFor
\newline
 \textbf{return} $\rho_T$
\end{algorithmic}
\end{algorithm}

We generated data sets of 2600 density matrices, 2000 of which formed our \defn{training set} whilst the rest formed our 
\defn{test set}.  The training set consisted of separable and PPT entangled states in equal part. NPPT entangled states were 
explicitly excluded from the training set since preliminary experiments showed that discriminating PPT entangled from separable states is harder than 
discriminating NPPT entangled from separable states.  Our test set contained states from all three classes in equal parts.
We also compared the performance of the different algorithms for training sets of increasing size, from 100 to 1000 density matrices, keeping the distribution of separable to PPT entangled states fixed at $50\%-50\%$.



\subsection{\label{sec:Implementation}Implementation of SVMs}

\subsubsection{Classical kernels}
We have tested the performance of SVMs with classical kernels considering an exhaustive search over a set of specified parameter values. The model selection incorporated Stratified K-fold cross validation with 5 folds and a shuffle of each class’ samples before splitting into batches. We have considered the following kernels:
    \begin{itemize}
        \item Polynomial: $\kappa(x,x')=\gamma \left< x,x' \right>^d$ with $d \in \left[ 2,7 \right] $
        \item Radial basis function (RBF): $\kappa(x,x')=\exp(-\gamma ||x -x'||^2)$
        \item Sigmoidal: $\kappa(x,x')=\tanh{(\gamma \left<x,x'\right>)}$
    \end{itemize}
    
 For all cases we checked the following values for the parameters: 
    \begin{align*}
       \gamma,\, C \in \lbrack &10^{-5}, 10^{-3.8889}, 10^{-2.7778}, 10^{-1.6667}, 10^{-0.5556}, 
        10^{0.5556}, 10^{1.6667}, 10^{2.7778}, 10^{3.8889}, 10^{5} \rbrack \\
    \end{align*}
where $\gamma$ is the parameter used in the kernel definition and $C$ is the regularization parameter as defined in Eq. \eqref{eq:SVM_lagDualForm}.

\subsubsection{Quantum-inspired kernels}
The quantum channel one uses to encode the classical alphabet $\cX$ onto the states of quantum mechanical systems strongly depends on 
the alphabet as well as the quantum resources one has available.  For instance basis encoding, rotation encoding, coherent state encoding 
\cite{schuld2021machine} and IQP \cite{havlivcek2019supervised} require a number of qubits much larger than the dimension of the alphabet.
Due to the high dimensionality of our dataset ($n^2 \times n^2$ for a $n \times n$ bipartite system), we opted to use amplitude encoding.  This encoding also allows for realistic simulation of our quantum inspired SVM on existing quantum devices with limited number of qubits  \cite{Rath2024}. 

In amplitude encoding each input vector ${\bf x}\in\cX$ is encoded using the complex amplitudes $\psi_i$ (see \eqnref{eq:vecexpansion}) 
of a quantum state relative to some orthonormal basis, henceforth taken to be the computational basis. Specifically, let 
$\mathbf{x}=(x_0,...,x_{N-1} ) \in \R^N $ be an input vector of dimension $N=2^n$. Then the amplitude encoding map is
described by the quantum channel $\Phi: \cX\to \cH_2^{\otimes n}$, such that $\Phi(\bf x)=\ketbra{\psi_{\mathbf{x}}}{\psi_{\mathbf{x}}}$ with
    \begin{equation}
        \ket{\psi_{\mathbf{x}}} = \frac{1}{\sqrt{\sum_{i}x_i^2}} \sum_{i=0}^{N-1} x_i \ket{i},
    \label{eq:amplitude_encoding}
    \end{equation}
where $\ket{i}$ denotes the i-th computational basis state. The denominator is a normalization factor to ensure that the output is a
valid quantum state. Note that for an alphabet of size $2^n$, amplitude encoding requires only $n$ qubits to efficiently store 
each of the feature vectors ${\bf x}$ making it a highly efficient feature map.

The quantum kernel associated to amplitude encoding is
    \begin{equation}
      \kappa(\mathbf{x},\mathbf{y})=
      \lvert\braket{\psi_{\mathbf{x}}}{\psi_{\mathbf{y}}}\rvert^2
      = (\mathbf{x}^T \mathbf{y})^{2}.
    \label{eq:amplitude_kernel}
    \end{equation}
    
Note that the quantum kernel is a polynomial kernel of degree 2~\cite{schuld2021machine} and as such can be efficiently computed using a classical computer. 
For bipartite states of small dimension, we evaluated this kernel running the quantum circuit described in section~\ref{sec:implRealQC}
using a quantum computer simulator and fed it to a classical optimizer for SVMs. However, for larger dimensions, we exploited the classical simulability of the kernel to directly evaluate it on a classical computer.

\section{\label{sec:results} Results}
\subsection {General considerations and experimental setup}

We have considered bipartite systems of dimension $3 \times 3$, $4\times 4$ and  $5 \times 5$, and generated their respective datasets formed by separable, PPT entangled and NPPT entangled quantum states as described in Sec.~\ref{sec:dataset}. For these datasets, we have performed a study of the prediction accuracy of the SVM fed with several classical and quantum-inspired kernels described in Sec.~\ref{sec:Implementation} with different sizes of the training dataset. The best performing classical kernels were determined by Grid Search among all the kernels and parameters specified in Sec. \ref{sec:Implementation}. 10 different grid searches were run for each system dimension and training set size. 

Working with the full dimensionality of the dataset is challenging due to the high number of attributes of the input data ($n^2 \times n^2$ for an $n \times n$ system) and the need of $N$ qubits such that $2^N \geq n^2 \times n^2$ to use amplitude encoding. We have used Principal Component Analysis (PCA) to reduce the dimensionality of the dataset and be able to reduce the number of qubits of the device. In the following sections, we have tested the impact of the choice of different number of components for the PCA that are used for the evaluation of the quantum-inspired kernel fed into the SVM. 

For coding the quantum circuits and the required computations, we have used the quantum computing framework \textbf{Pennylane} \cite{bergholm2018pennylane}. SVM training with classical or quantum-inspired kernels has been performed using \textbf{Scikit-learn}~\cite{scikit-learn} svm.SVC class and the grid search for the best classical models was implemented using the function GridSearchCV. The models have been run on Proteus supercomputing services of the Institute Carlos I for Computational Physics at the University of Granada \footnote{\url{https://proteus.ugr.es/}}.  


\subsection{Equivalence between kernels}
\begin{figure}[ht]
    \centering
    \subfloat[$3 \times 3$]{
        \includegraphics[width=0.65\textwidth]{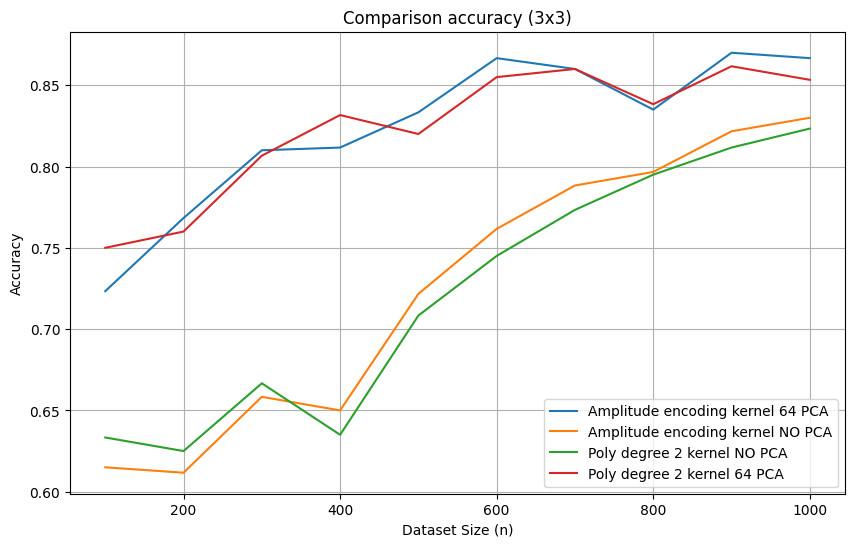}
        \label{fig:AEKvsPoly2K_3x3}
    }
    \hfill
    \subfloat[$4\times 4$]{
        \includegraphics[width=0.65\textwidth]{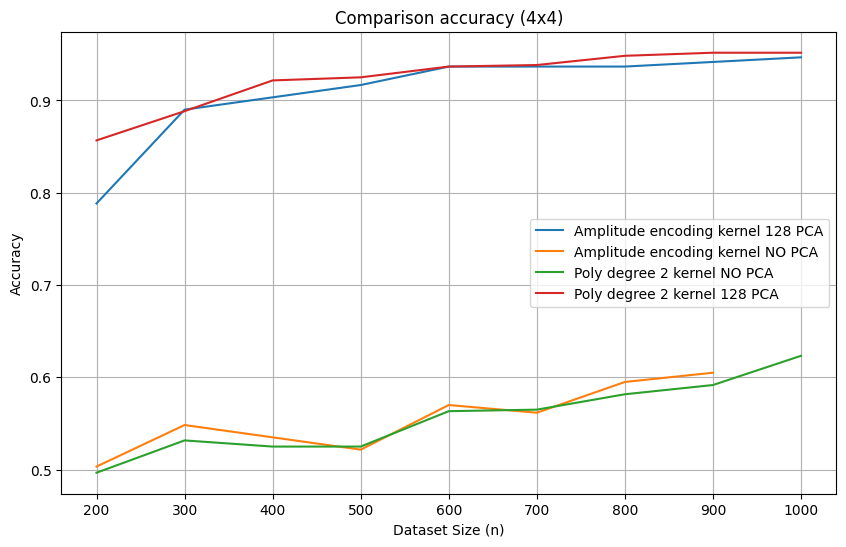}
        \label{fig:AEKvsPoly2K_4x4}
    }
    \caption{Accuracy comparison for a classical SVM trained with quantum-inspired amplitude embedding kernel and the polynomial kernel of degree 2. Dimensionality reduction with PCA was applied as indicated. Training and test sets were described in Sec. \ref{sec:methods}. }
    \label{fig:AEKvsPoly2K}
\end{figure}

Our first result was to verify that the amplitude encoding kernel and the classical polynomial kernel of degree 2 fed into a classical optimizer for SVMs provided equivalent results, as can be seen in figure~\ref{fig:AEKvsPoly2K}. 
Since the computation of the kernel matrix as the inner product of quantum states becomes computationally expensive with the number of qubits, we exploit the fact that the amplitude embedding kernel can be efficiently computed classically for the remaining results. 

\subsection{Impact of the PCA}
\begin{figure}[H]
\centering
\includegraphics[width=0.65\linewidth]{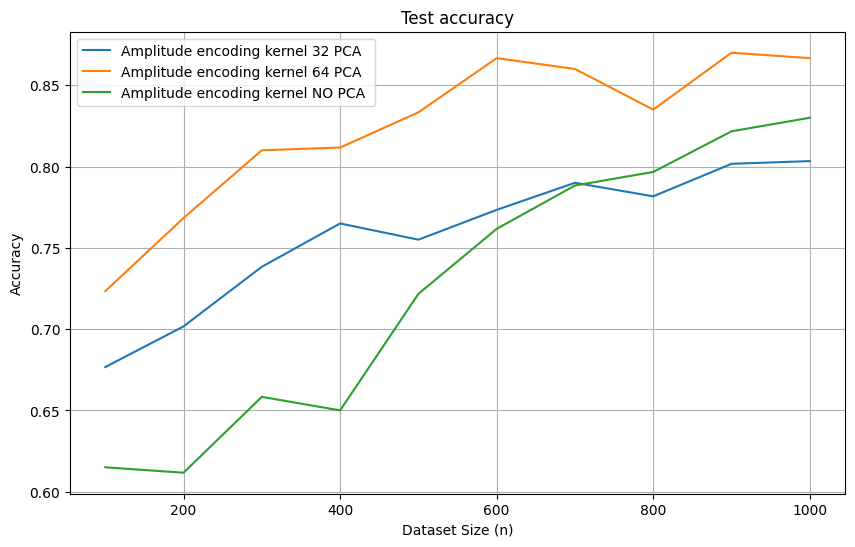}
\caption{\label{fig:3x3_pca}Comparison of the accuracy on the $3 \times 3$ system for the SVM with amplitude embedding for 5 (Quantum 32 PCA), 6 (Quantum 64 PCA) and 7 (Quantum NO PCA) qubits. }
\end{figure}

We have studied in more detail the impact of the dimensionality reduction with PCA on the results obtained with the quantum inspired amplitude embedding or polynomial kernel of degree 2 on the $3 \times 3$ system (see figure~\ref{fig:3x3_pca}).  

We have selected different number of components guided through the dimensionality of the data we could encode using amplitude encoding. Full-dimensionality $3 \times 3$ data required $81$ attributes. Therefore, for using a 5-qubit device, we applied PCA with $32$ ($ = 2^5$) components; for 6 qubits, PCA with $64$ ($ = 2^6$) components and since 7 qubits allowed us to encode $2^7=128$ attributes, we did not use PCA in that case. However, using a 7 qubit device for this problem forced us to use padding to perform amplitude encoding to a larger state space. 

The best performance was obtained for $6$ qubits in this case, probably because we maximize the amount of information in use from the original data without requiring the use of padding on a larger state space. Too much reduction, as shown with the 5 qubits representation, causes loss of information that affects the performance of the model. Therefore, in the experimentation that follows for the $3 \times 3$, $4 \times 4$ and $5 \times 5$ systems, we restricted the use of PCA to one qubit less than required for coding the full-dimensionality data, as summarized in Table~\ref{table:PCA}. 

\begin{table}[h]
    \centering
    \begin{tabular}{cccc}
        \toprule
        System Size & Matrix Size & Vector Size & Number of qubits used \\
        \midrule
        $3 \times 3$ & $9 \times 9$ & 81 & 6 \\
        $4 \times 4$ & $16 \times 16$ & 256 & 7 \\
        $5 \times 5$ & $25 \times 25$ & 625 & 9 \\
        \bottomrule
    \end{tabular}
    \caption{\label{table:PCA}Number of qubits used for coding the features of the different systems sizes according to the PCA strategy chosen. }
\end{table}

\subsection{Performance comparison}

\begin{figure*}[ht]
\centering
\includegraphics[width=0.65\linewidth]{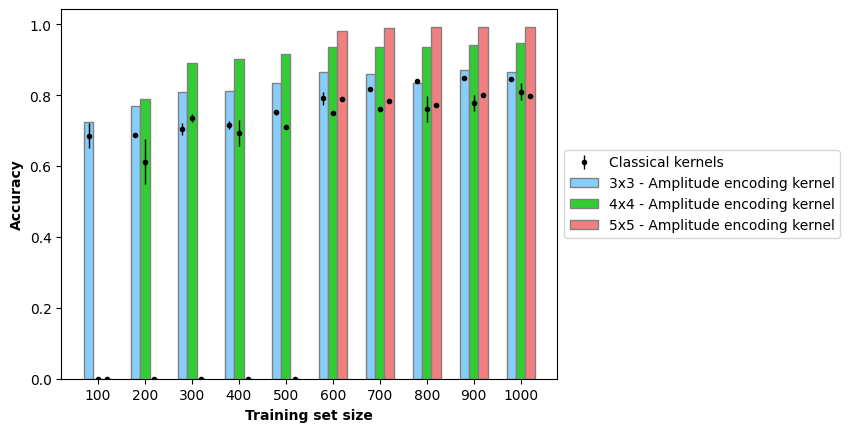}
\caption{Comparison of the accuracy for $3 \times 3$, $4 \times 4$ and $5\times 5$ systems. Bars indicate the best performing combination of PCA and quantum-inspired amplitude-encoding kernel for the different sizes of the problem. For the $5\times 5$ system we used a PCA for $2^9=512$ components which prevented us from applying this coding to a dataset with less than $512$ samples (the same applies to the $4\times 4$ system for the dataset of size 100). Dots and lines represent the average and standard deviation (if $\neq 0$) of best classical kernels resulted from 10 runs of GridSearch. }
\label{fig:all_sizes}
\end{figure*}

Our experiments show an increase of the performance with the size of the system for all the models under consideration (see \autoref{fig:all_sizes}), with average performances close to $0.8$, $0.9$ and $1.0$ for the $3\times3$, $4\times4$ and $5\times5$ systems, respectively. Regarding the best performing kernel in terms of accuracy, the quantum-inspired amplitude encoding kernel with PCA provided better performance than the rest of classical kernels that were considered for the GridSearch. 

\begin{figure}[H]
    \centering
    \subfloat[$3\times 3$]{%
        \begin{minipage}{0.65\textwidth}
            \centering
            \includegraphics[width=\linewidth]{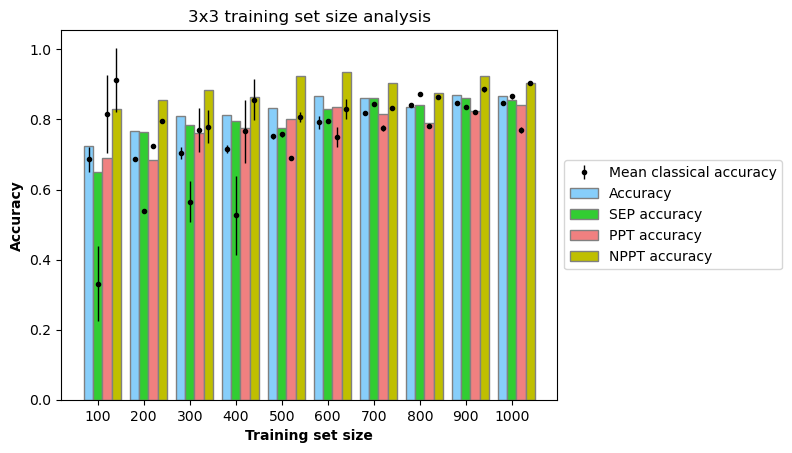}
        \end{minipage}
        \label{fig:subfig1}
    }
    \hfill
    \subfloat[$4\times 4$]{%
        \begin{minipage}{0.65\textwidth}
            \centering
            \includegraphics[width=\linewidth]{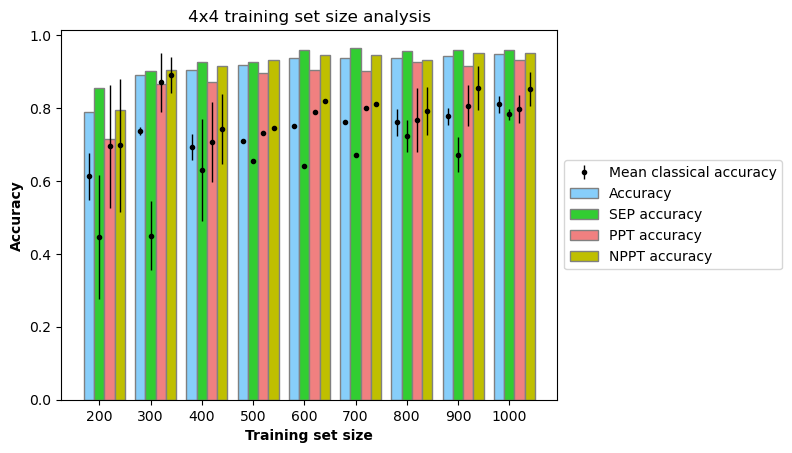}
        \end{minipage}
        \label{fig:subfig2}
    }

    \vspace{1em}

    \subfloat[$5\times 5$]{%
        \begin{minipage}{0.65\textwidth}
            \centering
            \includegraphics[width=\linewidth]{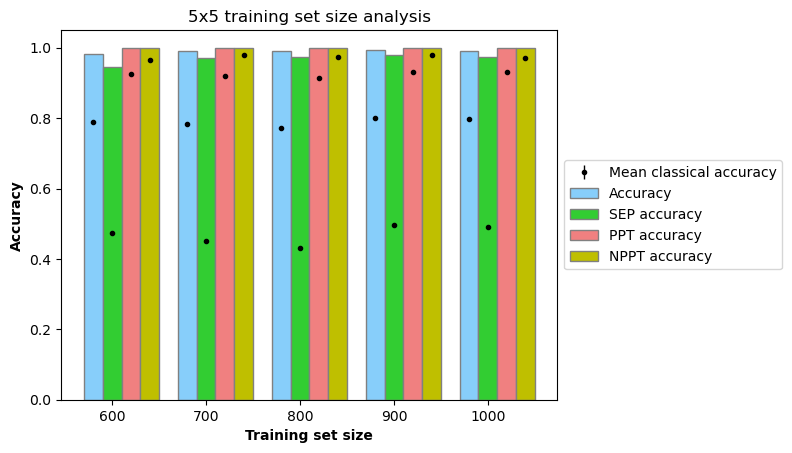}
        \end{minipage}
        \label{fig:subfig3}
    }

    \caption{Test accuracy for the average and SEP/PPT/NPPT classes of the SVM trained with different dataset sizes for the quantum-inspired amplitude encoding kernel with respect to the best performing classical kernels from Sec.~\ref{sec:Implementation}. Dots and lines represent the average and standard deviation (if $\neq 0$) of best classical kernels resulted from 10 runs of GridSearch.  }
    \label{fig:all_sizes_desglosada}
\end{figure}

Regarding the detection capability for different classes, \autoref{fig:all_sizes_desglosada} shows the test accuracy for the separable and PPT/NPPT-entangled classes. These plots show that the best performing classical kernel tends to provide biased results toward the entangled class as the size of the training datasets shrinks or as the dimension of the systems grows. However, the quantum-inspired kernel with PCA provided consistent results in all different classes for all training datasets and system sizes. 

\section{\label{sec:discussion} Discussion}

\subsection{Biases in data - Purity}
\begin{figure}[ht]
\centering
\includegraphics[width=0.48\linewidth]{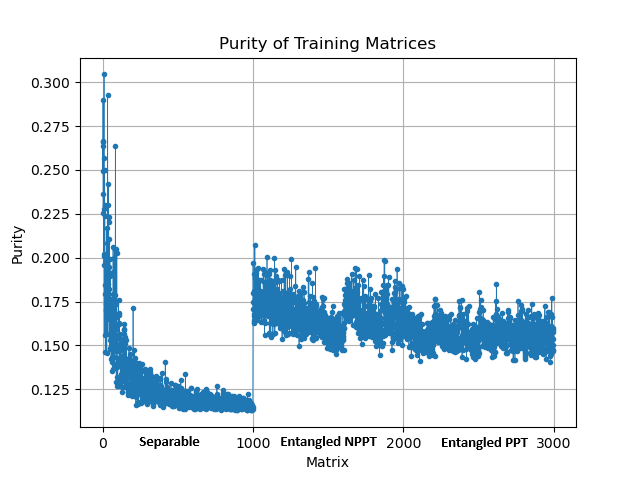}
\includegraphics[width=0.48\linewidth]{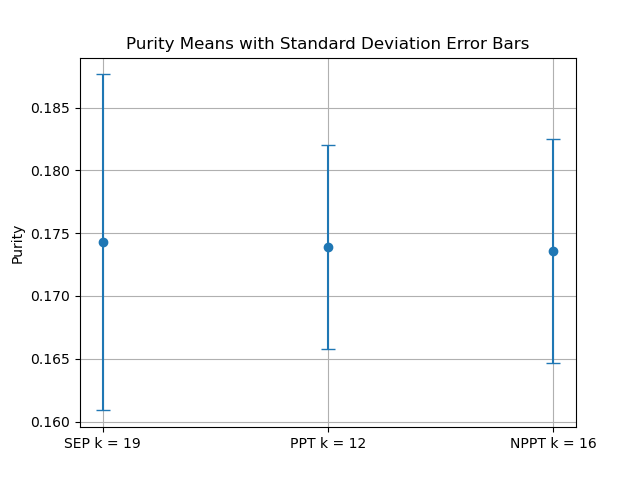}
\caption{Left: Plot of purity values of $3 \times 3$ dataset from paper \cite{casale2023large}. Right: $3 \times 3$ plot of means and standard deviation of the purity for entangled-PPT, entangled-NPPT and separable states for the dataset generated in this work. }
\label{fig:casale_purity}
\end{figure}

In order to be sure that our machine learning algorithm is indeed capable of discriminating entangled from non-entangled states 
we must ensure that the training set is free of any bias.  In our case such bias arises from the purity of the density matrices 
due to the way with which these density matrices are constructed.  As we explain in \secref{sec:dataset} the data set is 
constructed by randomly generating $k$ $d_Ad_B\times d_Ad_B$ matrices (either separable or entangled) and summing them up (see 
\eqnref{eq:sep_state}). The more matrices one uses in the construction of a $\rho\in\cB(\cH_a\otimes\cH_B)$ the lower its 
purity.  Indeed, it is known that $\rho\in\cB(\cH_a\otimes\cH_B)$ is separable if 
$\tr \left[\rho^2 \right]<(d_ad_B-1)^{-1}$~\cite{Gurvits2002}. Moreover, if the number of summands $k$ is different for separable states and  entangled states---resulting in different purities---we run the risk that our machine learning algorithm uses this bias to  classify the states, as opposed to identifying genuinely entangled states.  We observed such a purity bias in the data set from Ref. \cite{casale2023large} shown in \autoref{fig:casale_purity} (left). In order to avoid such a purity bias, we tuned the number of summands, $k$, such that the purity of 
$\rho\in\cB(\cH_A\otimes\cH_B)$ lies within a given range, for all three classes of states. Determining the mean and standard deviation of the purity for each class of states allowed us to select a homogeneous dataset for all three classes of states. The range of purities from which we sampled for the case $d_A=d_B=3$ is shown in \autoref{fig:casale_purity} (right).



\subsection{\label{sec:implRealQC}Implementation in real Quantum Computers}
In this section we discuss the resources required in order to implement quantum SVMs on real quantum hardware.  These resources are divided into two parts: (i) the 
resources needed in the amplitude encoding of the data and (ii) the resources needed to implement the relevant quantum kernel. In our scenarios of interest, a quantum SVM will work with quantum representations of our classical input data: i.e., quantum representations of training data and test data.

In this case, the classical input data is given to us in terms of the Bloch vector ${\bf r}\in\R^{d^2-1}$. This is a classical description of a quantum state comprising $d^2-1$ real numbers, which may be obtained by means of quantum state 
tomography~\cite{paris2004quantum}.   The memory needed to represent one input quantum state is thus $\cO(m d^2)$ bits, where $m =\cO(\mathrm{poly}(\log\delta^{-1}))$ is 
the classical memory needed to represent a real number in floating-point arithmetic with $\delta$ precision. 

A quantum SVM algorithm will be initialized by taking input Bloch vectors $\mathbf{x}$, $\mathbf{y}$ and loading them
into a quantum register of $\cO(\log(d))$ qubits using  amplitude encoding. By applying quantum synthesis algorithms such as the Solovay-
Kitaev algorithm~\cite{dawson2006solovay}, this step can be implemented via local quantum circuits $U_\mathbf{x}$, $U_\mathbf{y}$ 
comprised of single and two qubit gates. This circuit's size is of order $\cO(d^2 \log ^{c}(1/\varepsilon ))$ up to an additive error $\varepsilon$ in total time 
$\cO(d^2 \log ^{c}(1/\varepsilon ))$~\footnote{$\log^{c}$ is a polynomial of constant degree $c$.}. For instance, to achieve exponential 
precision $\varepsilon\sim \cO(1/2^m)$, the total cost scales as $\cO(d^2 m ^{c})$. Quantum computer algorithms often operate in the 
regime $\varepsilon\in\cO(1/\mathrm{poly}(n))$, which is where most quantum speedups occur~\cite{Nielsen_Chuang_2010}. 

\begin{figure}[ht]
    \centering
        \includegraphics[width= 1\textwidth]{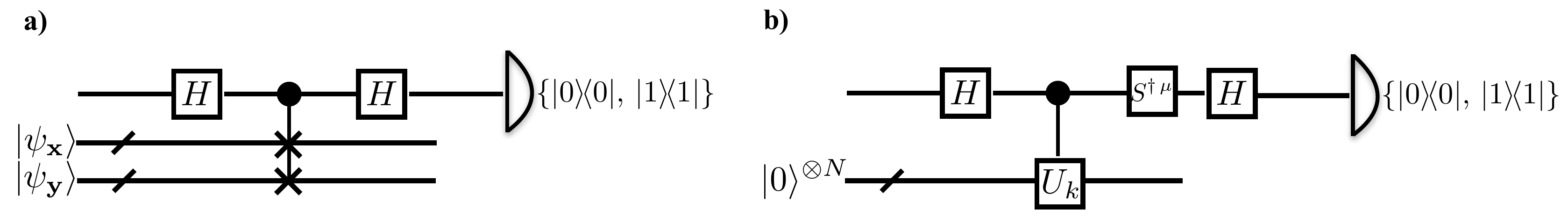}
    \caption{Quantum circuit implementations for calculating the quantum kernel of \eqnref{eq:amplitude_kernel}. a) SWAP test. b) Hadamard test. }
    \label{fig:SWAPHad}
\end{figure}

Once our classical data $\mathbf{x}$ is loaded into a quantum computer, we need to evaluate the kernel of our SVM using local quantum  circuits.  In our case, the kernel is nothing more than the squared overlap between two quantum states, which can be estimated using two  well-known quantum algorithms: the SWAP ~\cite{BuhrmanSWAP} or the Hadamard tests~\cite{aharonov2006}. Their  circuit  implementations are shown in \figref{fig:SWAPHad}. 

The SWAP test calculates the absolute value squared of the overlap between two  quantum states $\ket{\psi_{\bf x}},\, \ket{\psi_{\bf y}}$.  The circuit requires a two-dimensional auxiliary quantum system (originally prepared in the state $\ket{0}$).  The single qubit Hadamard gates are  given by $H=\frac{1}{\sqrt{2}}\bematrix 1 & 1 \\ 1 & -1\ematrix$.  The two qubit gate is the control swap gate: \[\proj{0}\otimes \one +\proj{1}\otimes\mathrm{SWAP}\] with $\mathrm{SWAP}\ket{\psi} \ket{\phi}=\ket{\phi}\ket{\psi}$.  The auxiliary qubit is measured in the computational basis with the probability of obtaining either of the two possible outcomes given by:\[p_k = \frac{1}{2}\left(1 +  (-1)^k\abs{\braket{\psi}{\phi}}^2\right), \, k\in\{0,1\}\]. 

The Hadamard test calculates the overlap $\braket{\psi}{\phi}\in\C$ between two quantum states $\ket{\psi_{\bf x}},\, \ket{\psi_{\bf y}}$.  Here the  multi-qubit register starts at the state $\ket{0}^{\otimes N}$ and the control unitary operator is given by: \[\proj{0}\otimes U_{0} + \proj{1}\otimes U_1\] where $U_0\ket{0}^{\otimes N}=\ket{\psi_{\bf x}}, \, U_1\ket{0}^{\otimes N}=\ket{\psi_{\bf y}}$.  When $\mu=0$, a measurement in the computational basis yields: \[p_k= \frac{1}{2}\left(1 + (-1)^k\mathrm{Re}\braket{\psi}{\phi}\right), \, k\in\{0,1\}\] whereas for $\mu=1$, one obtains: \[p_k=\frac{1}{2}\left(1 + (-1)^k\mathrm{Im}\braket{\psi}{\phi}\right), \, k\in\{0,1\}\] The unitary gate $S$ is given by $S = \bematrix 1 & 0 \\ 0 &-\ii\ematrix$.

As both the SWAP and the Hadamard tests yield similar circuit complexity, we will outline the resource costs  for the implementation of the Hadamard test.  This circuit operates on a total of $2\lceil \log_2(d))\rceil +1$ qubits: the two registers encoding the classical data plus a single auxiliary system as control. Each execution of the Hadamard test requires two controlled unitary operations, namely, the controlled versions of the unitaries  $U_{\bf x}, U_{\bf y}$ that are responsible for encoding the classical data. Using standard circuit  constructions~\cite{Nielsen_Chuang_2010}, these controlled unitary operations can be realized in the same circuit complexity as the state  preparation circuits described above. Additionally, we need at most three additional single-qubit gates (depending on whether we estimate  the real or imaginary part of the overlap).  

In  the alternative method, the SWAP test, these controlled-unitaries are replaced with a controlled SWAP, which is preceded, again, by the circuits which load the classical states into data. This variation does not fundamentally change the quantum algorithm, although it can be more convenient in scenarios where many states are to be prepared in a pre-processing step or in parallel.

To estimate the inner product with high confidence to within an additive error $\epsilon$, one must repeat the Hadamard test multiple times to gather statistics. The required number of runs can be 
determined using the Central Limit Theorem and Hoeffding's inequality to be $\nu = \cO(1/\epsilon^2)$. Hence the implementation of  a quantum SVM with a training set of size $M$ requires estimating 
all $\sim M^2/2$ pairwise inner products, which implies a worst case total runtime of $\cO(M^2/\epsilon^2)$ circuit executions. 

Thus the overall cost of implementing a quantum SVM given a training set of size $M$  of classical data ${\bf r}\in\R^{d^2-1}$ to 
$\epsilon$ precision is $\cO(d^2 \log ^{c}(1/\varepsilon)\times M^2/\epsilon^2)$.
As the dimension $d$ of the data scales exponentially with the number of quantum systems, it is clear that the cost of preparing the 
requisite quantum states can dominate the SVM training time.  This analysis shows that quantum SVMs are exponentially worse than their 
classical counterparts in the regime where exponentially precise kernel evaluations are required. For relatively small system sizes such 
as $d\in\{3, 4\}$ the costs are manageable but already for the case of $d=5$ we saw a clear advantage in computing the kernel using a classical 
SVM. We note, however, that a quantum advantage can in principle be obtained if we are given the 
classical training set in a quantum form, directly, i.e., if the state is experimentally available and we do not need to handle 
tomographic data. Advantages for quantum SVMs are also expected in settings where the error in the kernel matrix evaluation is inverse 
polynomial, rather than inverse exponential, which is the case for efficiently preparable entangled states such as those that arise from 
experimentally feasible scenarios~\cite{Huang22}.

Lastly, a clear advantage of the quantum setting, is the logarithmic savings in quantum memory that we obtain once the classical data is  loaded into a quantum computer. This may be advantageous in client-server scenarios in distributed computing, where a more powerful  server could do quantum data state preparations for a less powerful client with access to a small quantum computer. This could be useful  for leveraging the use of resources in distributed quantum computing. Quantum SVMs may also be of interest in scenarios where a server  would not want to hand in full classical data to a client because of privacy reasons (i.e., if medical data of real life patients is  being handled). Similarly to cryptographic protocols such as BB84 quantum key distribution or Wiesner's quantum money  \cite{Nielsen_Chuang_2010}, once the classical data is loaded into a quantum form, a classical eavesdropper would destroy the quantum  superposition upon measurement. Beside, they would typically need an exponential number of samples in order to recover the encoded classical data reliably. A full analysis of the cryptographic security of such protocols is beyond the scope of this work.

\section{Conclusions}

In this work, we developed and analyzed support vector machine (SVM)-based algorithms for detecting entanglement in quantum systems of dimensions $3\times 3$, $4\times 4$, and $5 \times 5$, where the positive partial transpose (PPT) criterion alone cannot provide a complete solution. Our approach successfully handles three categories: separable states, entangled states detectable by PPT, and entangled states undetectable by PPT. The results demonstrated that SVMs equipped with a quantum-inspired kernel are highly effective for entanglement detection, particularly for states beyond the reach of the PPT criterion. Notably, the performance improved significantly with system dimension, achieving accuracies of $80\%$, $90\%$, and nearly $100\%$ for $3\times 3$, $4\times 4$, and $5 \times 5$ systems, respectively. This suggests that higher-dimensional systems, which exhibit more complex entanglement features, may benefit even more from machine learning approaches compared to traditional methods.

Our study also revealed the crucial role of PCA in handling limited datasets. When working with only a few hundred states, dimensionality reduction to 64 principal components already yielded strong results, while SVMs without PCA performed substantially worse. This finding highlights PCA's importance for practical applications where data may be scarce or computationally expensive to obtain.

A critical challenge identified in our work relates to biases in data generation, particularly concerning state purity. Since entanglement is inherently linked to purity, the choice of algorithm for generating mixed states significantly impacts the resulting dataset and, consequently, the model's performance. This purity bias must be carefully considered to ensure reliable and generalizable entanglement detection. 

Finally, while our study focused on classical simulations, we examined the prospects for implementing these methods on real quantum hardware. Current limitations, such as noise and gate errors, pose challenges for direct deployment, but emerging error-mitigation techniques may provide viable solutions in the near future.

Looking ahead, several promising directions emerge from this research. Hybrid quantum-classical SVMs leveraging quantum kernel methods could potentially enhance detection capabilities, especially for larger systems. Developing more robust data generation techniques that minimize purity-related biases will be essential for creating reliable benchmarks. Additionally, exploring the scalability of these methods to systems beyond $5 \times 5$ dimensions remains an important open question. Our results collectively demonstrate that machine learning, particularly when combined with quantum-inspired approaches, offers a powerful and flexible framework for entanglement detection that complements traditional criteria. As quantum technologies continue to advance, addressing the challenges of data quality and hardware limitations will be crucial for translating these theoretical advances into practical applications.

\section{Acknowledgments}

We want to acknowledge funding from Ministry for Digital Transformation and of Civil Service of the Spanish Government through projects, QUANTUM ENIA project call - Quantum Spain project, and by the European Union through the Recovery, Transformation and Resilience Plan - NextGenerationEU within the framework of the Digital Spain 2026 Agenda; PID2021-128970OA-I00 10.13039/501100011033 funded by MCIN/AEI/10.13039/501100011033/ "FEDER Una manera de hacer Europa". Also the FEDER/Junta de Andalucía program A.FQM.752.UGR20. MS acknowledges support from Ayuda Ramón y Cajal 2021 (RYC2021-032032-I, MICIU/AEI/10.13039/501100011033, ESF+) as well as Project FEDER C-EXP-256-UGR23 Consejería de Universidad, Investigación e Innovación y UE Programa FEDER Andalucía 2021-2027. JBV acknowledges support from Ayuda Ramón y Cajal 2022 (RYC2022-036209-I, MICIU/AEI/10.13039/501100011033, ESF+), Ayuda Consolidación (CNS2023-145392, MICIU/AEI/10.13039/501100011033, NextGenerationEU/PRTR) and Horizon Europe FoQaCiA (GA 101070558).
Finally, we are also grateful for the technical support provided by PROTEUS, the supercomputing center of the Institute Carlos I for Theoretical and Computational Physics in Granada, Spain.


\bibliographystyle{sn-mathphys-num}


\end{document}